  \documentclass[]{aastex}
  \usepackage{emulateapj5,onecolfloat,graphicx}
  \usepackage[hyperindex,breaklinks]{hyperref}

  \def\eg{{\it e.g.},}
  \def\etal{{\it et al.}}

\def\jcoph{J. Comp.\ Phys.}

\def\etc{{\it etc.}}
\def\ie{{\it i.e.},}

\def\DF{{\small DF}}

\def\pmb#1{\setbox0=\hbox{$#1$}%
  \kern-0.25em\copy0\kern-\wd0
  \kern.05em\copy0\kern-\wd0
  \kern-0.025em\raise.0433em\box0}
\long\def\Ignore#1{\relax}

\def\ILR{{\small ILR}}
\def\OLR{{\small OLR}}

\def\WASER{{\small WASER}}

\slugcomment{To appear in ApJ}

\begin{document}

\twocolumn[
\title{Spiral Instabilities in $N$-body Simulations I: Emergence from Noise}
\author{J. A. Sellwood}
\affil{Department of Physics \& Astronomy, Rutgers University, \\
       136 Frelinghuysen Road, Piscataway, NJ 08854-8019, USA \\
       {\it sellwood@physics.rutgers.edu}}

\vspace{0.25cm}

\begin{abstract}
The origin of spiral patterns in galaxies is still not fully
understood.  Similar features also develop readily in $N$-body
simulations of isolated cool, collisionless disks, yet even here the
mechanism has yet to be explained.  In this series of papers, I
present a detailed study of the origin of spiral activity in
simulations in the hope that the mechanism that causes the patterns is
also responsible for some of these features galaxies. \par In this
first paper, I use a suite of highly idealized simulations of a
linearly stable disk that employ increasing numbers of particles.
While the amplitudes of initial non-axisymmetric features scale as the
inverse square-root of the number of particles employed, the final
amplitude of the patterns is independent of the particle number.  I
find that the amplitudes of non-axisymmetric disturbances grow in two
distinct phases: slow growth occurs when the relative overdensity is
below $\sim 2\%$, but above this level the amplitude rises more
rapidly.  I show that all features, even of very low amplitude,
scatter particles at the inner Lindblad resonance, changing the
distribution of particles in the disk in such a way as to foster
continued growth.  Stronger scattering by larger amplitude waves
provokes a vigorous instability that is a true linear mode of the
modified disk.
\end{abstract}


\vspace{0.25cm}
]

\section{Introduction}
Despite considerable effort over many decades, no theory for the
origin of spiral patterns in galaxies has yet gained wide acceptance.
There is good evidence \citep[\eg][]{KN79,KKC11} that some patterns in
galaxies are excited by tidal interactions
\citep[\eg][]{BH92,SL93,Dobb10}, others may be driven by bars
\citep[\eg][]{Salo10}, and perhaps even dark matter halo substructure
\citep[\eg][]{Dubi08}.  But when all these agents are excluded from
$N$-body simulations, they still manifest spiral patterns and it
therefore seems reasonable to expect that self-excited features also
develop in galaxies.

Several reviews \citep{Toom77,Atha84,BL96,BT08,Sell10a} give detailed
summaries of theoretical efforts to account for self-excited spirals.
Most theorists agree that spirals are gravitationally-driven density
waves in the old stellar disk, a point of view that is strongly
supported by the light distribution in the near IR
\citep[\eg][]{GPP04,ZCR9,KKC11} and by streaming motions in high
spatial resolution velocity maps \citep[\eg][]{Viss78,Chem06,SVOT}.
However, strong disagreements arise over other aspects; in particular,
the lifetimes of the spirals.

Small-amplitude, non-axisymmetric Jeans instabilities of equilibrium
models are of two fundamental types.  Cavity modes are standing waves,
reminiscent of those in organ pipes and guitar strings, in which
traveling waves reflect off an inner boundary, often the galaxy
center, and corotation, where they are amplified.  The bar-forming
instability is the strongest mode of many simple models
\citep[\eg][]{Kaln78,Jala07} and is of the cavity type \citep{Toom81}.
The other type of mode, such as the edge mode \citep{Toom81,PL89} or
groove mode \citep{LH78,SK91}, is driven from corotation.  All these
are vigorous instabilities that exponentiate on an orbital time scale.

\citet{BL96} argue that spirals are caused by a more gentle
\WASER\ instability \citep{Mark77}, a cavity mode that reflects off a
``$Q$-barrier'' instead of the center, that may become quasi-steady at
finite amplitude through non-linear effects.  However, their specific
examples \citep{BLLT} of low-mass, cool disks that are needed to
support slowly-growing bi-symmetric modes have been shown
\citep{Sell11} to develop more vigorous multi-arm instabilities that
quickly alter the basic state they invoke.

\cite{GLB65} and \citet{Toom90} argue that, in galaxies lacking a bar
or a recent interaction with a companion, spirals are no more than
shearing bits and pieces of swing-amplified noise.  Substantial
inhomogeneities, such as star clusters and giant molecular clouds,
raise the noise level far above that expected from the stars alone,
and the noise level can be amplified $\sim 100$-fold in cool,
responsive disks.  However, since the spirals they wish to account for
themselves cause the disk to heat quickly, it is hard to understand
how a galaxy could sustain the combination of responsiveness and noise
level needed to create spiral features of the amplitude we observe.
Furthermore, this mechanism has difficulty accounting for even a
modest degree of symmetry in the patterns \citep[\eg][]{KKC11}.

From the earliest days \citep{MPQ71,HB74,JS78}, $N$-body simulations
of dynamically cool collisionless disks have manifested, recurrent
short-lived transient spirals.  Claims of long-lived waves in some
simulations are not reproducible \citep{Sell11}.  \citet{SC84} showed
that spiral activity in purely stellar disks is self-limiting, because
potential fluctuations caused by the rapidly evolving spirals
themselves scatter stars away from circular orbits; the increased
random motion in the disk reduces its ability to support further
collective oscillations.  \citet{SC84} also found that spiral activity
can continue ``indefinitely'' in models that mimic the effects of gas
dissipation and star formation, offering an attractive explanation
for the observed coincidence between spiral activity and gas content
in galaxies.  The behavior they reported continues to occur in modern
simulations of galaxy formation \citep[\eg][]{Abad03,Rosk08,Ager11}.

Local theory \citep{JT66} predicts that a perturbing mass moving on a
circular orbit in a self-gravitating disk surrounds itself with a
wake.  In an $N$-body simulation, each particle is itself a perturber
and has its own peculiar motion that is typically as large as that of
the surrounding particles whose motion it affects.  Note that a system
of dressed particles, with each orbiting at its own angular rate, will
give rise to shearing density fluctuations that will be strongest when
the wakes are aligned, giving the appearance of swing amplification.
In fact, the picture of randomly placed dressed particles in a
shearing disk is equivalent to that of swing-amplified noise, with the
input noise level raised by the mass cloud surrounding each source
particle \citep{Toom90}.

\citet{TK91} studied swing-amplified particle noise using simulations
in shearing $(x,y)$-coordinates in a periodic box.  They had to
counter the continuous rise in the level of random motion in order to
be able to study the long-term equilibrium level of spiral activity,
and they therefore added a ``dashpot'' type damping term to the radial
acceleration, $f_x = -C v_x$.  The level of random motion should have
settled to a roughly constant value if the damping constant $C$ were
proportional to the inverse of the particle number.  They found,
however, that the observed rms velocity of the particles exceeded
their predicted level, because the density fluctuations were about
twice as large as they expected -- a discrepancy that they attributed
to additional correlations between the particles that developed over a
long period.

While the origin of spiral features in global simulations has yet to
be explained, their amplitude in larger $N$-body simulations seems to
be too great to be just swing-amplified particle noise \citep{SC84}.
\citet{Fuji11} found that spirals in simulations with larger numbers
of particles heated the disk more slowly, allowing the spiral activity
to continue for longer.  These authors incorrectly attributed the
rapid heating rate in the smaller $N$ simulations by \citet{SC84} to
two-body relaxation, whereas the higher level of shot noise simply
provokes stronger spirals.\footnote{If collisional relaxation were
  rapid, it would have been impossible for the same code to have
  supported the global modes \citep{SA86,ES95,SE01} predicted by
  linear perturbation theory for a collisionless stellar fluid, or to
  match the results from direct integration of the collisionless
  Boltzmann equation \citep{INS84}.  The formula for the relaxation
  rate in a two-dimensional disk of particles does not contain the
  Coulomb logarithm \citep{Rybi72} and the rate is therefore
  suppressed more effectively by gravity softening.}

In this paper, I attempt to uncover the reason for the suprisingly
vigorous spiral activity in $N$-body simulations of cool,
collisionless disks.  The models employed here are highly idealized in
order to simplify the dynamics to the point that it becomes
understandable.  Improved realism is deferred to later papers.

I have long hoped \citep{Sell91,Sell00} that, in realistic galaxy
models, the decay of one spiral feature might change the background
disk in such a manner as to create conditions for a new instability,
much as \citet{SL89} had observed in simulations of a cool, low-mass
mass disk having a near-Keplerian rotation curve.  Here I finally
present the evidence that something similar to this idea really does
occur in more galaxy-like models, but the behavior differs in detail
from that I envisioned.

\section{A Linearly Stable Disk}
The disk in which the circular speed is independent of radius, now
known as the ``Mestel disk'' \citep{BT08}, is predicted by linear
stability analyses \citep{Zang76,Toom81,ER98} to be globally stable to
most non-axisymmetric disturbances.  I therefore use this model in the
simulations described in this paper.

The potential in the disk plane is
\begin{equation}
\Phi(R) = V_0^2 \ln \left( {R \over R_i} \right),
\end{equation}
where $V_0$ is the circular speed, and $R_i$ is some reference radius.
The surface density of the full-mass disk is
\begin{equation}
\Sigma(R) = {V_0^2 \over 2\pi G R}.
\end{equation}
A star at radius $R$ moving in the plane with velocity components
$v_R$ and $v_\phi$ in the radial and azimuthal directions respectively,
has specific energy $E = \Phi(R) + (v_R^2 + v_\phi^2)/2$ and specific
angular momentum $L_z = Rv_\phi$.  A distribution function (\DF) for
this disk that yields a Gaussian distribution of radial velocities of
width $\sigma_R$ is \citep{Toom77,BT08}
\begin{equation}
f(E,L) = \cases{ F L_z^q e^{-E/\sigma_R^2} & $L_z>0$ \cr 0 & otherwise, \cr}
\end{equation}
where $q = V_0^2/\sigma_R^2 - 1$ and the normalization constant
\begin{equation}
F = {V_0^2 / (2\pi G) \over 2^{q/2}\surd\pi (q/2-0.5)!\sigma_R^{q+2}}.
\end{equation}

In order to avoid including stars having very high orbital frequencies
near the center, \cite{Zang76} introduced a central cut out in the
active surface density by multiplying the distribution function $f$ by
the taper function
\begin{equation}
T_i(L_z) = {L_z^\nu \over (R_iV_0)^\nu + L_z^\nu} ,
\end{equation}
where the index $\nu$ determines the sharpness of the taper.  This
taper function removes low angular momentum stars, reducing the active
disk mass at the center, which must be replaced by unresponsive matter
in order that the total potential is unchanged.  The characteristic
radius $R_i$ of this cut out introduces a scale length.  Henceforth, I
adopt units such that $V_0 = R_i = G = 1$.

\cite{Zang76} found that this almost full-mass disk was stable to all
bisymmetric modes, provided the index $\nu \leq 2$.  This important
result was confirmed by \cite{ER98}, who also found that other disks
with power-law rotation curves were remarkably stable to bisymmetric
modes.  We now understand that this disk model is linearly stable to
bar-forming modes through Toomre's (1981) mechanism, because the
frequency singularity at the disk center ensures that disturbances of
any pattern speed $\Omega_p$ will have an inner Lindblad resonance
(hereafter \ILR) as long as $m>1$.

However, Zang also found that the full-mass Mestel disk was globally
unstable to lop-sided ($m=1$) modes, as are the power-law disks
\citep{ER98}.  This instability persists no matter how hot the disk or
gentle the central cut out.  The mechanism for the lop-sided mode is
also a cavity mode, since feedback through the center is allowed only
for $m=1$ waves because the \ILR\ condition in power-law disks cannot
be satisfied when $\Omega_p>0$.

Swing-amplification in a disk having a flat rotation curve is most
vigorous for $1 \la X \la 2.5$ \citep{JT66,Toom81}, where $X = Rk_{\rm
  crit}/m$, with the smallest wavenumber for axisymmetric
instabilities $k_{\rm crit} = \kappa^2/(2\pi G\Sigma)$
\citep{Toom64,BT08}.  Since $X=2/m$ in a full-mass Mestel disk, the
swing-amplifier is at its most vigorous for $m=1$ disturbances.  The
parameter $X = 2/(\xi m)$ is increased by reducing the active surface
density of the disk to $\xi\Sigma$ without changing the circular
speed; here $0 \leq \xi \leq 1$.  Since the swing-amplifier is all-but
dead when $X>3$ in a flat rotation curve, the lop-sided instability
should disappear for $\xi \la 2/3$.

\cite{Toom81} reports that a half-mass ($\xi=0.5$) disk, with $\nu=4$
for the central cut out, has no small-amplitude, global instabilities
for any $m$.  This model is the only known globally stable disk with
differential rotation and significant disk mass.\footnote{Two of the
  composite $\Omega$ models reported by \cite{Kaln72} appear to be the
  only other known stable disks, but are of less interest because they
  rotate uniformly.}

As the disk still extends indefinitely to large $R$, \cite{Toom88}
added an outer taper
\begin{equation}
T_o(L_z) = \left[ 1 + \left( {L_z \over R_oV_0} \right)^\mu \right]^{-1}, 
\end{equation}
with index $\mu$ to control the abruptness of the taper about the
angular momentum characteristic of a circular orbit at some large
radius $R_o$.  Again, the potential is unchanged if the mass removed
is replaced by rigid matter.  He found $\mu \leq 6$ to be adequate to
avoid provoking outer edge modes \citep{Toom81,PL89}.

\begin{figure}[t]
\includegraphics[width=\hsize]{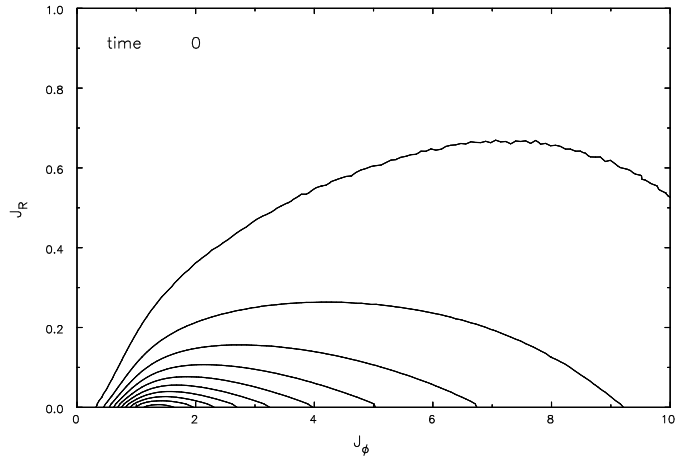}
\caption{Initial density of particles in the space of the two actions.
  The contour spacings are linear from 5\% to 95\% of the function
  maximum.  The density for low values of $J_\phi$ is suppressed by
  the inner angular momentum taper and, as is generally true, the
  density is greatest for near circular orbits ($J_R=0$) and decreases
  for more eccentric orbits.}
\label{act.1.0}
\end{figure}

As this taper function still does not limit the radial extent of the
disk, but merely reduces the active density of the outer disk, I have
to add the additional constraint that $f(E,L_z) = 0$ when $E >
\Phi(R_{\rm max})$, so that the disk contains no particles with orbits
that extend beyond $R_{\rm max}$.  In order that this additional limit
does not also provoke an edge mode, I choose $R_{\rm max} \ga 1.7R_o$,
so that it removes mass only where the disk density is already very
low.

In the simulations reported in this paper, I adopt $\xi=0.5$,
$q=11.44$ so that the half-mass disk has nominal $Q=1.5$, the inner
cut out is characterized by $\nu=4$ at $R_i=1$, and the outer taper
has $\mu=5$ at $R_o=11.5$ and the limiting radius is $R_{\rm max}=20$.

Figure~\ref{act.1.0} shows the density of particles in the space of
the two actions $J_R$ and $J_\phi$ \citep{BT08}.  In an axisymmetric
two-dimensional disk model, the azimuthal action $J_\phi \equiv L_z$,
while the radial action,
\begin{equation}
J_R = {1 \over \pi} \int_{R_{\rm min}}^{R_{\rm max}} v_R \; dR,
\end{equation}
expresses the amount of radial motion the particle possesses.  Here
$v_R = \left[ 2E - 2\Phi(R) - L_z^2/ R^2 \right]^{1/2}$ (positive root
only in this definition) and the limits of integration are the radii
of peri- and apo-center of the orbit where the argument of the square
root is equal to zero.  Thus $J_R$ also has the dimensions of angular
momentum and $J_R=0$ for a circular orbit.  The actions are an
alternative set of integrals to the classical integrals $E$ and $L_z$,
but the relation between the two sets of integrals is cumbersome.

\section{Simulations of the ``Stable'' Disk}
The model just described, with the outer taper but not the energy
cut-off, was predicted by \cite{Toom88} to have no small-amplitude
instabilities.  Here I report simulations with this model both to test
the prediction and to study how replacing the infinitely finely
divided ``stellar fluid'' by a finite number of particles affects the
behavior.

I use the two-dimensional polar-grid code \citep{Sell81} with multiple
time step zones \citep{Sell85}, that has previously been shown to
reproduce the predicted normal modes of unstable disks
\citep{SA86,ES95,SE01}.  Gravitational forces between particles obey a
Plummer softening law, with the softening parameter $\epsilon =
R_i/8$, so that forces can be thought of arising from a disk of
thickness $\epsilon$.

The grid has 128 spokes and 106 radial rings, and I restrict
disturbance forces to the $m=2$ sectoral harmonic only; the
axisymmetric radial force is $-V_0^2/R$ at all radii.  I employ a
basic time step of $\delta t = R_i/(40V_0)$, which is increased by
successive factors of 2 for particles with radii $R>R_i$, $R>2R_i$,
$R>4R_i$, \& $R>8R_i$.  Timesteps are also sub-divided when particles
move near the center in a series of ``guard zones'' \cite{SS04},
without redetermining the gravitational field, which is an adequate
approximation since the dominant force in this region arises from the
rigid component.  Direct tests showed that the results are insensitive
to moderate changes of grid size or time step.

\begin{figure}[t]
\includegraphics[width=\hsize]{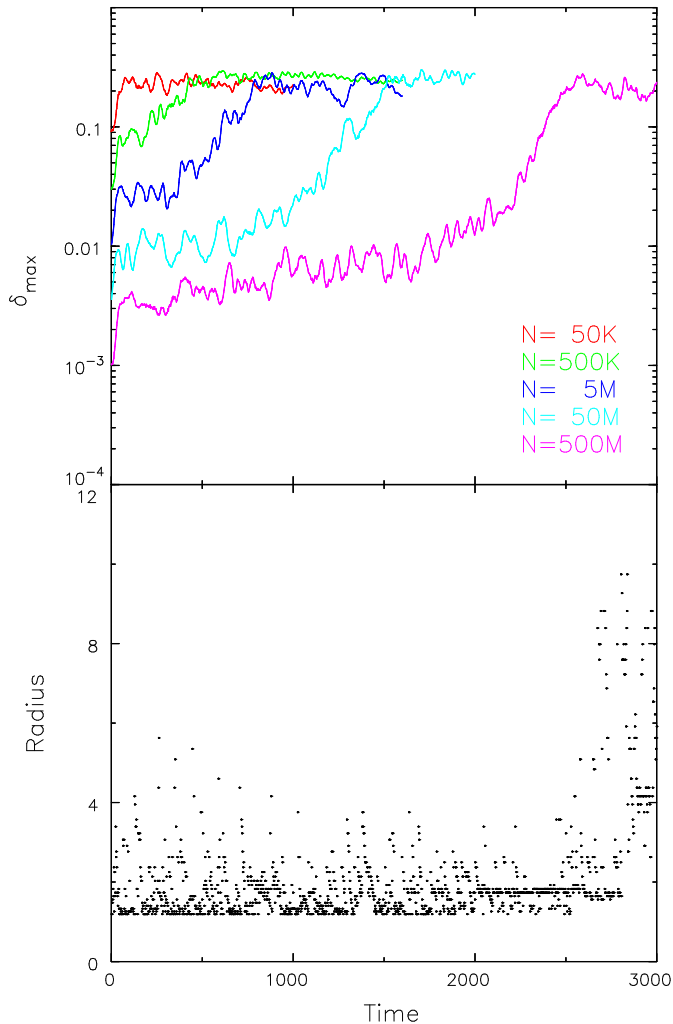}
\caption{Upper panel: Time evolution of the peak overdensity in a
  series of simulations of the half-mass Mestel disk with different
  numbers of particles.  The model is predicted by \citet{Toom88} to
  be globally stable.  The ordinate reports the maximum value of
  $\delta\Sigma/\Sigma$ on grid rings over the range $1.2 < R < 12$
  where the surface density is little affected by the tapers.  Linear
  theory predicts the amplitude should remain proportional to
  $N^{-1/2}$. \hfil\break Lower panel shows the radius of the
  measurement reported in the upper panel for run 500M only.  The
  apparent lower bound is because values for $R < 1.2$ were excluded
  to eliminate shot noise in the tapered part of the disk.}
\label{cntrst}
\end{figure}

This paper reports results from several simulations that differ in the
number of particles, and I label each run by the number of particles.
Much of the analysis focuses on run 50M, which therefore has $5 \times
10^7$ particles, and its variants runs 50Ma, 50Mb, \etc

\section{Results}
Figure~\ref{cntrst} shows the evolution of the relative overdensity of
bi-symmetric disturbances in a series of simulations with increasing
numbers of particles.  The upper panel shows the quantity $\delta_{\rm
  max}$, which is the largest value of the ratio $\delta \Sigma /
\Sigma$ over the radial range $1.2 < R < 12$, \ie\ the part of the
disk where the surface density has not been significantly reduced by
the tapers.  The unit of time is $R_i/V_0$ and, since I have chosen
units such that $R_i=V_0=1$, the orbit period at radius $R$ is
$2\pi R$.

As more particles are employed, the initial value of $\delta_{\rm
  max}$ decreases as $N^{-1/2}$, as it must for randomly placed
particles.  Its value rises by a factor of a few in the first few
dynamical times as the initial shot noise is swing-amplified.  In the
smallest $N$ experiment, the initially amplified noise is already
close to the saturation level, but as $N$ rises, the amplitude takes
increasingly long to reach its maximum.  However, the final amplitude
is independent of $N$ for all particle numbers up to $N=5\times10^8$.
In the larger $N$ experiments, the amplitude increase is characterized
by roughly exponential growth at two distinct rates: an initial period
of slow growth, followed by a steeper rise once $\delta_{\rm max} \ga
0.02$.  These two separate rates of growth are approximately
independent of $N$ over the amplitude ranges where they are observed.

It might seem that exponential growth indicates an unstable normal
mode, or perhaps a few such instabilities, but such an interpretation
is unattractive for a number of reasons.  Simulations of linearly
unstable models \citep[\eg][]{SA86,ES95} usually reveal the mostly
vigorous instability emerging from the noise at an early stage and
dominating the subsequent growth until it saturates.  Modulated growth
could occur in a system that supports a small number of
overstabilities having similar growth rates but different pattern
speeds, since the changing relative phases of the modes over time
causes beat-like behavior.  In all such cases, however, the
overstabilities should maintain phase coherence until they saturate,
but power spectra of these simulations (shown below) reveal multiple
features none of which retains phase coherence throughout the period
of growth.  Furthermore, were the later, more rapid rise due to the
emergence of more rapidly growing normal modes of the original disk,
it is unusual that they should take so long to rise above the
amplitude of the more slowly growing ``modes'' that would also need to
be invoked to account for the initial slow rise, and it is most
unlikely that the change of slope would occur at almost the same
amplitude, as suggested for the three larger $N$ experiments
(Figure~\ref{cntrst}).  Finally, Toomre's (1981) linear stability
analysis that predicted the model to be globally stable also argues
against this interpretation of the results.

\begin{figure*}[p]
\centerline{\includegraphics[width=.75\hsize]{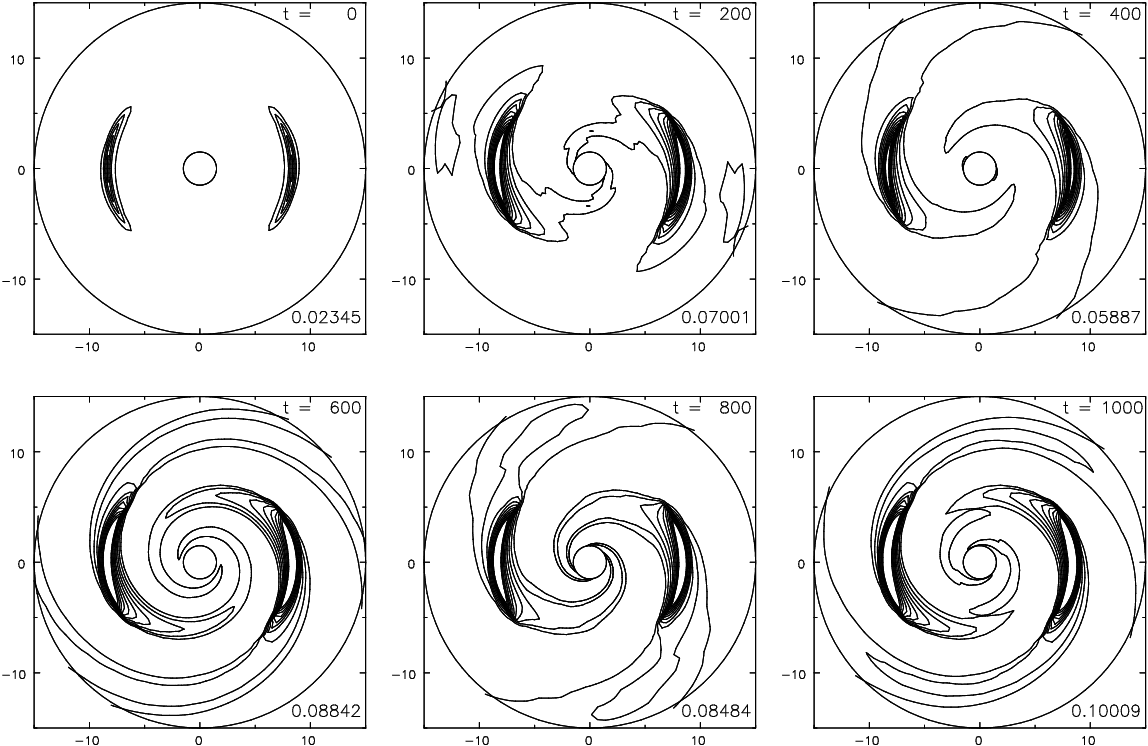}}
\caption{Gradual development of polarization in the mass
  distribution of run 50M.  The contours show the positive part only
  of the disturbance density, with the bisymmetry enforced by the
  code, and contour levels are the same in each panel.  The inner
  boundary is at $R=1.5$, the outer at $R=15$, and the ``source''
  particles are at $R=7$.  The peak relative overdensity at each time
  is reported in the lower right corner.}
\label{contdn}
\bigskip
\centerline{\includegraphics[width=.52\hsize,angle=270]{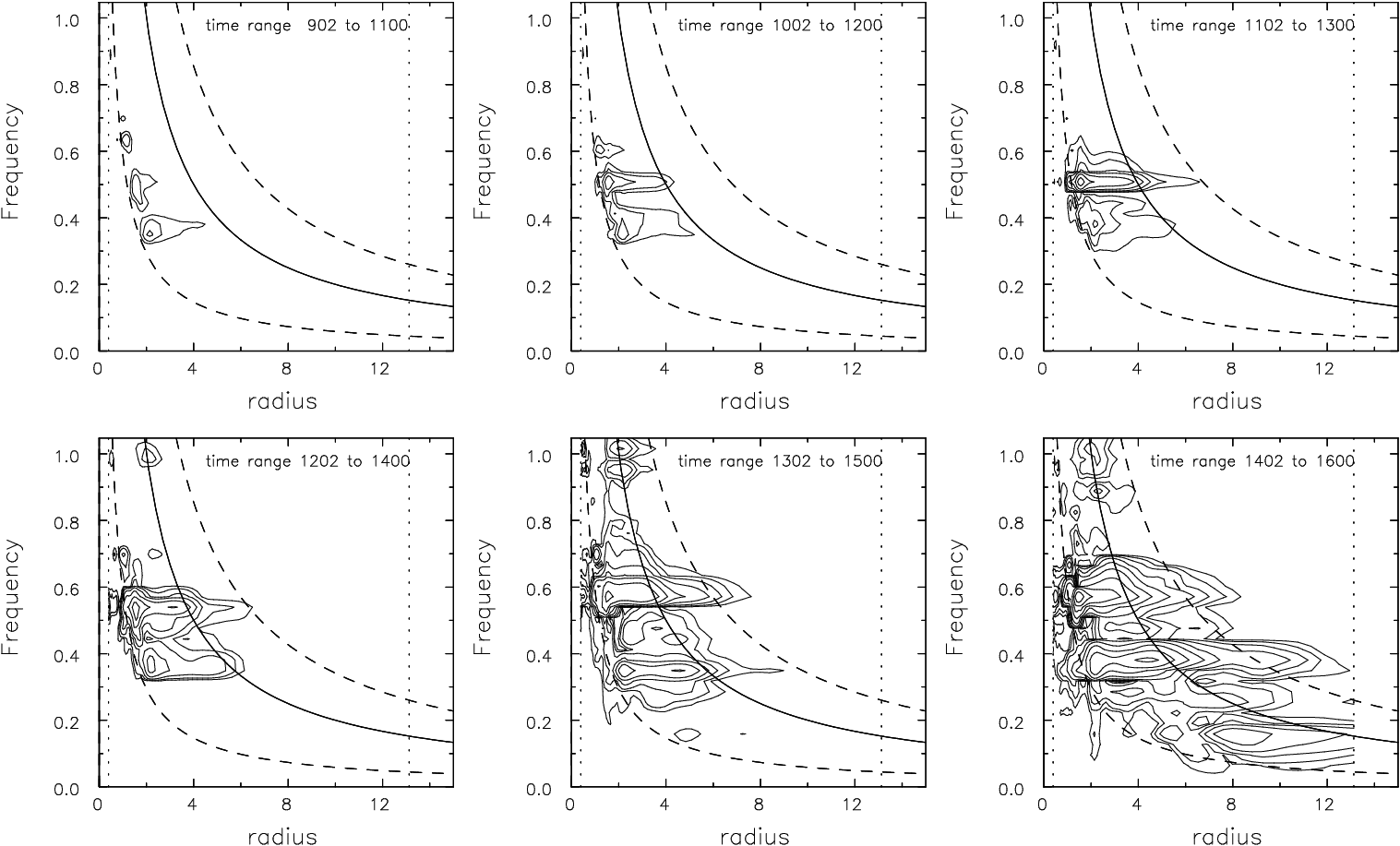}}
\caption{Power spectra of $m=2$ disturbances in run 50M.  Successive
  contour levels, which start from the same value in each plot,
  differ by factors of two.  Each panel is taken from data over the
  period indicated, which overlap so that only half the plots are from
  independent data.  The solid curves show $2\Omega(R)$ and the dashed
  curves $2\Omega(R) \pm \kappa(R)$.  Data outside the radial ranges
  marked by vertical dotted lines, where the surface density is low,
  are excluded because they are too noisy.}
\label{spctrm}
\end{figure*}

\subsection{Slow growth}
\label{slowrise}
The lower panel of Figure~\ref{cntrst} shows that the maximum
disturbance density in run 500M is generally in the inner disk for $t
< 2500$, \ie\ until the saturation amplitude is reached.  The behavior
over the entire early period $0 < t < 2000$ seems to follow a
quasi-repetitive pattern: a density maximum appears in the range $3
\la R \la 6$ that then propagates inwards.  This pattern results from
swing-amplified noise creating a trailing spiral disturbance that
travels inwards at the group velocity, as shown in the ``dust to
ashes'' Figure of \citet{Toom81}.  \citet[][their Figure 7]{SL00}
report very similar behavior in their three-dimensional simulations of
a low-mass exponential disk embedded in a rigid halo.

The gradual rise of spiral activity in the global simulations reported
here can also be viewed as the build-up of mass clouds surrounding
each particle.  Figure~\ref{contdn} shows that each particle can be
regarded as being ``dressed'' by a spiral wake, as in the local
context.  To construct this Figure at each instant, I stacked copies
of the grid-estimated density distribution from run 50M by scaling the
entire grid radially and rotating it such that the density maximum on
each ``source'' ring in turn lay at $R=7$ and zero azimuth.  At the
initial moment, the non-axisymmetric part of the combined density
shows no features other than the azimuthal extension of the source
caused by forcing $\cos2\theta$ angular dependence.  As time
progresses, each source particle becomes dressed by a wake of
gradually increasing mass and spatial scale; the maximum over-density
shows a general rise, with fluctuations that closely follow the time
variations of $\delta_{\rm max}$ in the same simulation, shown by the
cyan line in Figure~\ref{cntrst}.

However, two related aspects of the behavior are especially
noteworthy.  First, growth in the large $N$ simulations continues for
a much longer period than expected.  \citet{JT66} found, from a local
linear analysis, that the wake takes $\sim 5$ epicycle periods to
become fully developed.  Swing-amplification of particle shot noise
causes the immediate rise in the first {\it few\/} time units, as
already noted, which is a large part of the development of spiral
wakes.  The epicyclic period at radius $R$ in the Mestel disk is
$2^{1/2}\pi R/V_0$, or $\sim 44$ time units at $R=10$, about halfway
out in the disk.  Thus the period of continued gradual growth greatly
exceeds five epicycle periods at a typical radius, which is the
timescale expected from linear theory.  Second, the simulations do not
reveal a limiting amplitude; instead more rapid growth takes over in
every case, even when $5 \times 10^8$ particles are employed.
Section~\ref{enhanced} presents further analysis of the slow-rise
phase that accounts for these differences.

\subsection{Rapid rise}
\label{rapidrise}
The acceleration in the rate of growth once $\delta_{\rm max} \ga
0.02$ is clearly inconsistent with the linear theory prediction.  The
enhanced growth rate is again approximately independent of $N$, as is
the final amplitude at a relative overdensity between 20\% and 30\%.

Figure~\ref{spctrm} shows a number of power spectra, \ie\ contours of
power as functions of radius and frequency, from run 50M.  A
horizontal ridge in these figures indicates a coherent density
disturbance that extends over a range of radii and has an angular
frequency $m\Omega_p$, with $m=2$ and $\Omega_p$ being the rotation
rate or pattern speed.  There is very little power at frequencies
higher than those shown.  The solid line indicates the frequency of
circular motion, $m\Omega$ and the dashed lines $m\Omega\pm\kappa$.
Each panel gives the spectrum over an interval of 200 time units, with
the start times of successive panels shifted forward by 100 time
units.

As shown in the lower panel of Figure~\ref{cntrst} for run 500M, density
fluctuations at early stages in run 50M also arise at radii $4 \la R
\la 7$ from swing-amplified shot noise and subsequently propagate
inward at the group velocity \citep{Toom69}.  The amplitude of each
disturbance rises as it approaches the \ILR\ because wave action is
conserved as the group-velocity decreases until wave-particle
interactions absorb it near the \ILR\ \citep{LBK72,Mark74}.

The behavior changes around $t \sim 1300$.  The strongest feature up
to this time has a frequency $\sim 0.5$ (\eg\ third panel of
Figure~\ref{spctrm}), but thereafter a new much stronger disturbance
appears with a frequency $\sim 0.55$.  Analysis with the mode fitting
apparatus described in \citet{SA86} using data from the period $1100
\leq t \leq 1500$ reveals that this wave has a frequency $\omega =
m\Omega_p + i\gamma \simeq 0.55 + 0.13i$, although the co-existence of
other waves over the same period makes this frequency estimate
somewhat uncertain.  The significant growth rate suggests it is a true
instability, yet it is scarcely conceivable that this mode could have
been growing at this rate from $t=0$, since that would require its
initial amplitude to have been smaller by $\ga 50\;e$-folds before it
first became detectable.

\begin{figure}[t]
\includegraphics[width=\hsize]{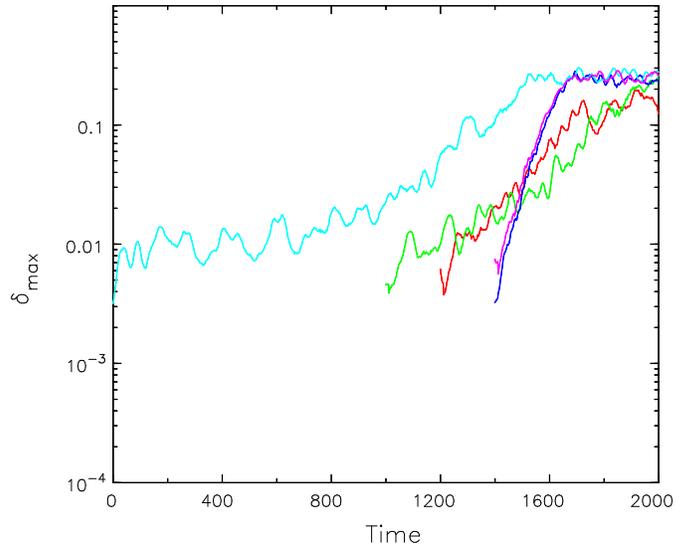}
\caption{Time evolution of the peak overdensity in five simulations
  with $N=50$M.  The cyan curve is for run 50M, reproduced from
  Figure~\ref{cntrst}, while the green, red, and magenta curves show
  the behavior of runs 50Ma, 50Mb, and 50Mc created by randomizing the
  azimuths of the particles at $t=1000$, 1200 and 1400 respectively.
  For the blue curve, run 50Md, the particles preserved the same $E$
  and $L_z$, but the radial phase was also reset.  Both runs restarted
  from $t=1400$ manifest a single dominant instability.}
\label{cntrst2}
\end{figure}

\subsection{Randomized restarts}
\label{restarts}
Figure~\ref{cntrst2} reports the results of four further simulations
that were restarted from rearrangements of the particle distribution
at selected times from run 50M.  The original run is reproduced as the
cyan curve.

To construct three of these simulations, I simply added to the
azimuthal phase, a random angle selected uniformly from 0 to
$2\pi$, while leaving the radius and velocity components in
cylindrical polar coordinates unchanged.  This procedure erases any
non-axisymmetric feature above the shot noise level, and the starting
amplitude is similar to that of the cyan line at $t=0$.

The green, red and magenta curves respectively show the evolution of
runs 50Ma, 50Mb, and 50Mc, which were re-started from $t=1000$,
$t=1200$, and $t=1400$.  In all cases the peak non-axisymmetric
density rises more quickly than that of the original run over the
period $200 \la t \la 1000$.  Furthermore, the later the restart time,
the more rapidly the amplitude rises and the {\it sooner\/} the
amplitude saturates!  The different behavior of these randomized runs
can only be a consequence of the changes to the particle distribution
that occurred in the original run prior to azimuthal randomization.

\begin{figure}[t]
\centerline{\includegraphics[width=.8\hsize]{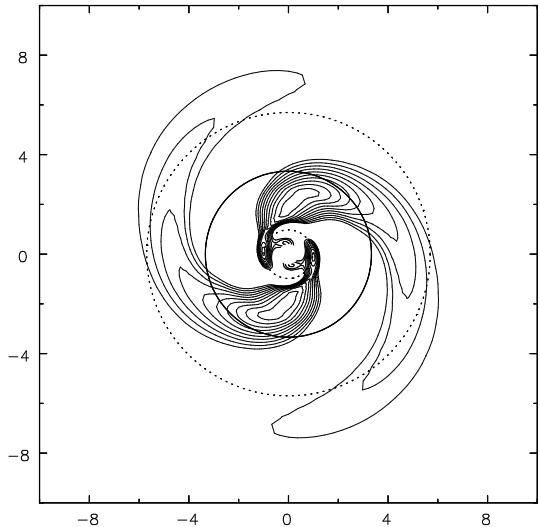}}
\caption{Best fit mode extracted from the period of exponential growth
  in run 50Mc using the procedure described by \citet{SA86}.  The
  circles mark the radii of the principal resonances.}
\label{mode}
\end{figure}

The most dramatically different behavior is that of run 50Mc
(magenta), which manifests a single vigorous instability of
eigenfrequency $\omega = 0.6+0.16i$ -- close to that of the dominant
disturbance around $t=1400$ identified in the earlier analysis of run
50M (Section~\ref{rapidrise}).  The form of the unstable mode is illustrated
in Fig~\ref{mode}.  Clearly, the particle distribution at $t=1400$ has
a feature that provokes this instability, and which must have been
created by the earlier evolution.

The behavior of the other two cases (green and red curves of
Figure~\ref{cntrst2}) differs not quite so markedly from the original,
but again indicates that the particle distribution at both $t=1000$
and $t=1200$ contains features that lead to an accelerated rise of
non-axisymmetric features.  I return to these cases in
Section~\ref{enhanced}.

Note that randomizing the azimuthal phases at $t=1400$ only made the
identifiable instability of the original model stand out more clearly.
Thus, the destabilizing agent cannot be a previously-created
non-axisymmetric feature, \eg\ a weak bar, in the particle
distribution, and must therefore be a feature in the distribution
either of the radial phases, or of the integrals ($E$ and $L_z$ or
$J_R$ and $J_\phi$).

In order to test whether changes to the distribution of radial phases
are important, I determined $E$ and $L_z$ for each particle and then
selected at random a new pair of radial and azimuthal phases from an
orbit having these integrals in the analytic potential of the Mestel
disk.  These new phases determine new positions and velocities for
each particle, although it has the same integrals as before.  The
evolution in this case, run 50Md shown by the blue line in
Figure~\ref{cntrst2}, makes it clear that the same instability is
present as that in run 50Mc (magenta line).  Thus the important change
by $t=1400$ of the original run is to the distribution of the
integrals.

\begin{figure}[t]
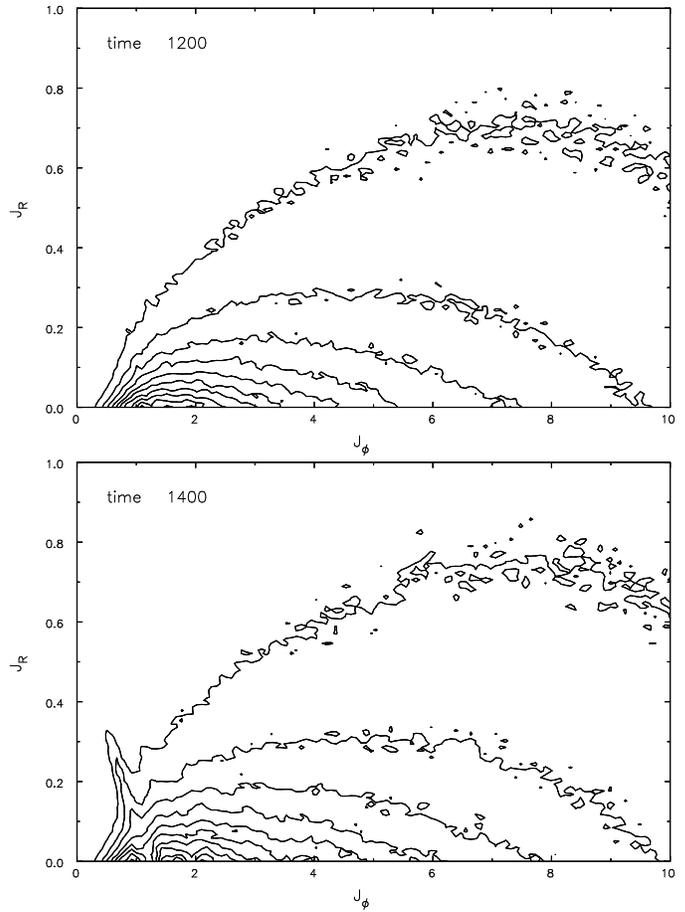

\includegraphics[width=\hsize]{act.1.1200.ps}
\smallskip
\includegraphics[width=\hsize]{act.1.1400.ps}
\caption{Action-space density of particles in run 50M at $t=1200$
  (upper) and $t=1400$ (lower).}
\label{act.change}
\end{figure}

\vfill\eject
\subsection{Changes to the distribution function}
\label{dfchanges}
\begin{figure}[t]
\includegraphics[width=\hsize]{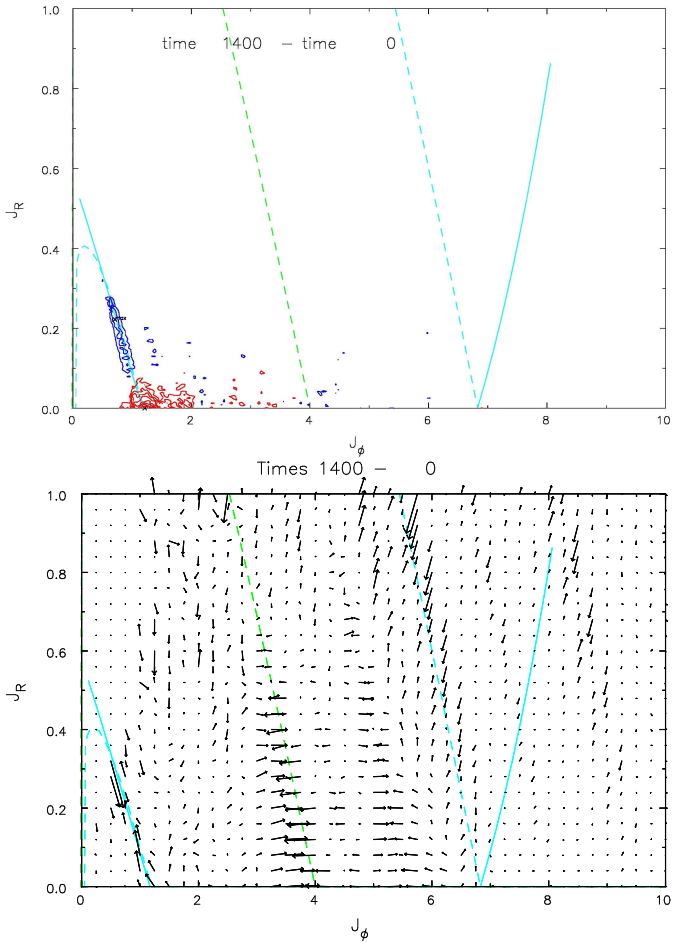}
\smallskip
\includegraphics[width=\hsize]{actv1400-0.ps}
\caption{Upper panel contours the differences between the action-space
  density of particles in run 50M at $t=1400$ and $t=0$, with positive
  (negative) differences being indicated by blue (red) contours.  The
  dashed lines show the loci of Lindblad resonances (cyan) and of
  corotation (green) for $\Omega_p=0.25$, while the solid lines
  indicate the scattering trajectory from $J_R=0$ for the Lindblad
  resonances. \hfil\break lower panel shows the mean displacements of
  particles in the same space and over the same period.  The arrow
  size is proportional to the mean value of the displacement in each
  element.}
\label{act.diff}
\end{figure}

Figure~\ref{act.change} shows the distribution of particles in action
space of run 50M at the times $t=1200$ (upper) and $t=1400$ (lower).
Although the contours are noisier, caused by a degradation of the
original smooth arrangement, their overall shape in the upper panel is
little changed from that at $t=0$ (Figure~\ref{act.1.0}).  However, more
substantial changes are visible in the lower panel.  The changes are
shown more clearly in the upper panel of Figure~\ref{act.diff}, which
contours differences between the distributions at $t=1400$ and at
$t=0$, with increases shown by the blue contours and decreases by red.

The various lines in Figure~\ref{act.diff} show resonance loci and
scattering trajectories for a disturbance of angular frequency
$m\Omega_p = 0.5$, which is the frequency of the dominant feature of
the power spectrum shown in the third panel of Figure~\ref{spctrm}.  The
resonance lines (dashed) are computed for $\Omega_p = \Omega_\phi +
l\Omega_R/m$, where $l=0$ for corotation and $l=\mp1$ for the inner
and outer Lindblad resonances, and the frequencies $\Omega_\phi$ and
$\Omega_R$ are computed for orbits of arbitrary eccentricity.  The
scattering trajectories (solid lines) are computed assuming $\Delta E
= \Omega_p\Delta L_z$, as required by the conservation of the Jacobi
constant in a rotating non-axisymmetric potential \cite{BT08}, with
$\Delta E$ converted to $\Delta J_R$ and assuming $J_R = 0$ initially
for a particle in resonance.  Note that any scattering at corotation
would occur with $\Delta J_R=0$.

It is clear that the principal changes to the distribution of
particles in action space were caused by scattering at the \ILR\ of
the strongest non-axisymmetric feature to have developed before
$t=1400$.  This conclusion is considerably strengthened by the
evidence in the lower panel of Figure~\ref{act.diff}, which shows the
mean displacements of particles in action space between $t=0$ and $t=
1400$.  This Figure not only confirms the convergence of vectors
towards the density maximum created by scattering at the \ILR, but
also reveals that substantial changes have taken place both at
corotation and at the \OLR\ that produce no net change in density.
Note that the vectors along the Lindblad resonance lines are parallel
to the scattering directions for those resonances, computed for the
same pattern speed, while they are near horizontal at corotation.
Thus particles at all three resonances are scattered with no change to
Jacobi's invariant for the measured pattern speed.  The displacements
at the \OLR\ cause no apparent net change to the density (upper
panel), while those at the \ILR\ do.  The different behavior at the
two Lindblad resonances is partly caused by the geometric consequence
of the ``focusing'' of waves that propagate inwards, while outgoing
waves spread out over an increasing disk area.  It is also significant
that the scattering direction at the \ILR\ is very close to the
resonance locus, so that particles stay on resonance as they lose
angular momentum, whereas a small gain or loss of angular momentum at
the \OLR\ moves the particle off resonance.

There are a few additional features visible in the lower panel of
Figure~\ref{act.diff} that appear to be due to another somewhat lower
frequency wave that is also present at $t=1400$ (see the fourth panel
of Figure~\ref{spctrm}).  These features are much weaker in an
equivalent plot (not included here) of the lower panel of
Figure~\ref{act.diff} constructed for $t=1300$.

\begin{figure}[t]
\includegraphics[width=\hsize]{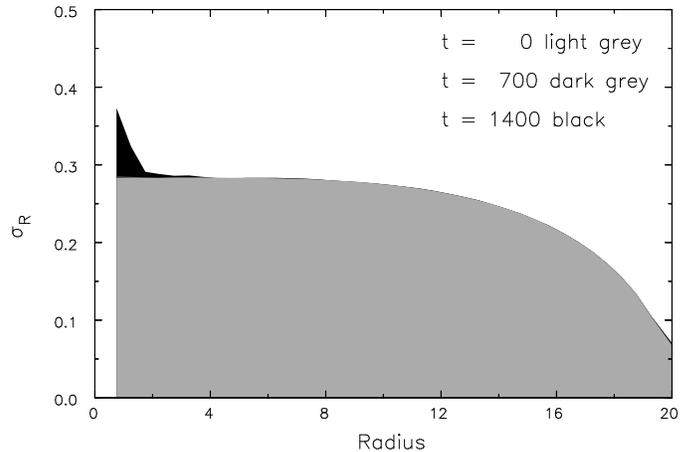}
\caption{Second moment of the radial velocity distribution of
  particles in run 50M at the three indicated moments.  No visible
  changes have occurred by $t=700$, while a significant increase can
  be seen near the center at $t=1400$.}
\label{sigRplt}
\end{figure}

Scattering at the \ILR\ naturally causes a very localized increase in
the random motions of the particles.  Figure~\ref{sigRplt} shows the
radial variation of the second moment of the radial velocities
estimated from the particles.  By construction from the analytic \DF\,
the radial velocity distribution is near Gaussian at $t=0$, but the
resonant scattering in action space does not preserve this property.
Note that this very localized heating at a resonance was caused by a
disturbance that had a relative overdensity of $\sim 2$\%.

\subsection{Nature of the new instability}
\label{newmode}
What causes the new instability that appears in runs 50Mc and 50Md,
and which is also present, but less distinct, in the continued run
50M?  Since randomizing the azimuthal coordinates makes the
instability stand out more clearly, it cannot be caused by trapping of
particles into some non-axisymmetric feature, neither can it be caused
by the kind of non-linear wave coupling discussed by \citet{MT97} or
\citet{Fuch05}.  Its rapid exponential growth characterizes it as a
vigorous, global, linear mode of the modified disk.

It has long been my expectation \citep[\eg][]{Sell91,Sell00} that the
recurrent cycle would prove to be one of groove-type modes
\citep{SK91} caused by depopulating the \DF\ near the \OLR\ of one
wave to provoke a new instability with corotation near the previous
\OLR.  Indeed this is what \citet{SL89} observed in their simulations
of a low mass disk with a near Keplerian rotation curve.  Yet I have
been unable to find evidence to support this expectation in more
massive disks with flatter rotation curves, and the new evidence
presented in Figure~\ref{act.diff} seems quite emphatically to rule it
out, at least for this case.  While changes to the \DF\ are clearly
responsible, it is not scattering at the \OLR, and some other mechanism
is at work.

The only significant change to have occurred to cause runs 50Mc and
50Md to behave so differently from the initial behavior of run 50M is
the localized heating near $R=1.17$, which is the radius of the
\ILR\ for near circular orbits for the scattering wave.  There are no
net changes elsewhere, even though the lower panel of
Figure~\ref{act.diff} indicates that some particles in other parts of
the disk have changed places.

Since the new instability has a slightly higher pattern speed than
that of the preceding scattering wave, the radii of the principal
resonances are all correspondingly smaller, and therefore the \ILR\ of
the new instability would lie within the heated region of the disk.
The localized deficiency of nearly circular orbits created by the
previous changes to the \DF\ will have strongly inhibited a supporting
response in this region of the disk to forcing by an inward
propagating density wave.

Thus it seems to me that the most plausible unstable mode created by
these ``initial'' conditions is a cavity-type mode that is a standing
wave between corotation and a hard inner reflection off the region of
the disk where the \DF\ has been changed.  Since the inner disk is
``hotter'', there is a superficial resemblance between this suggestion
and the \WASER\ mechanism proposed by \citet{Mark77} and advocated by
\citet{BL96}.  However, the \WASER\ mode has a low growth rate because
the out- and in-going waves at corotation are both trailing, resulting
in very mild amplification as the disturbance connects only the long-
and short-wave branches of the dispersion relation.  Furthermore,
their mode requires a low surface mass density and $Q \sim 1.0$ near
corotation, whereas this model has a massive disk with $Q=1.5$.
Finally, the change in disk properties that causes the inner
reflection is quite different from the simple increase in $Q$
postulated for the \WASER\ mechanism.  Much more vigorous modes, such
as that observed here, seem likely to require a full swing from
leading to trailing \citep{Toom81}.  Thus I suspect that the in-going
trailing wave reflects off the region where the \DF\ has been changed
as an outgoing leading wave.  I have no evidence that a hard
reflection can occur in this case, so this proposed mechanism is
entire speculation.

\begin{figure}[t]
\centerline{\includegraphics[width=.98\hsize]{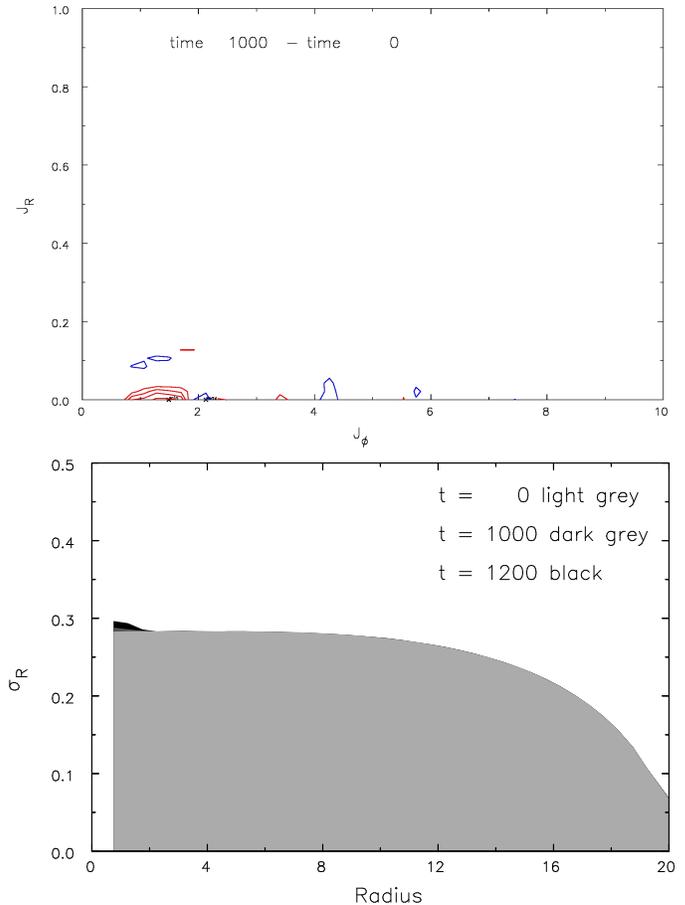}}
\smallskip
\includegraphics[width=\hsize]{sigRplt2.ps}
\caption{Upper panel contours the differences between the action-space
  density of particles in run 50M at $t=1000$ and $t=0$, with positive
  (negative) differences being indicated by blue (red) contours.  The
  kernel width is five times larger than that used in
  Figure~\ref{act.diff} and the contour levels are one
  fifth. \hfil\break Lower panel shows the radial velocity dispersion
  of particles in run 50M at the three indicated moments.  The changes
  from $t=0$ are non-zero only near the center; they are tiny by
  $t=1000$ and moderately larger at $t=1200$.}
\label{sigRplt2}
\end{figure}

\subsection{Earlier restarts}
\label{enhanced}
Simulations 50Ma and 50Mb were restarted at $t=1\,000$ and $t=1\,200$
respectively.  As shown in Figure~\ref{cntrst2}, density fluctuations
in these runs grew at rates that were intermediate between that of the
original run 50M and the clear instability exhibited by those (50Mc
and 50Md) that were restarted at $t=1\,400$.  Moreover, these two runs
support multiple disturbances having several different angular
frequencies, none of which maintains coherence through to the
saturation amplitude.  The randomized particle distributions at these
intermediate times must also differ in important ways from the smooth
particle distribution of run 50M at $t=0$.

Figure~\ref{sigRplt2} shows that the changes to the \DF\ by $t=1000$ are
tiny, but nevertheless significant.  I had to reduce the contour
levels in order to reveal the tiny changes to the density of particles
in action space (top panel), which necessitated a proportionate
increase in the kernel size so that the shot noise from the number of
particles contributing to the lowest contour is unchanged.  It is
clear that there has been some significant scattering of particles
away from the $J_R=0$ axis, especially at small radii.  These, and the
significantly larger changes that occur by $t=1200$ (not shown) caused
some heating of the innermost part of the disk, as illustrated in the
lower panel.  This very mild heating of the inner disk is caused by
the absorption at the \ILR s of incoming disturbances created by
swing-amplified shot noise at larger radii.  Although the wave
amplitudes were tiny, they caused a lasting and significant change to
the \DF.

The randomized restarts at $t=1000$ (run 50Ma) and $t=1200$ (run 50Mb)
confirm that these tiny changes, and not non-linear coupling say, were
responsible for the enhanced growth of non-axisymmetric structure.  It
seems likely that the small changes that have taken place are
sufficient to cause some partial reflection of waves incident on the
inner disk that boosts the density fluctuations, but is apparently
insufficient to provoke an indefinitely growing, more vigorous mode at
these times.

Thus the slow rise in the density fluctuations described in
Section~\ref{slowrise} occurred because the physical system gradually
develops a marginally modified inner disk.  Mild scattering of
particles in the inner disk as the wave action of the noise-driven
structures created at larger radii is absorbed at the appropriate
\ILR.  These changes to the dynamical properties of the inner disk
seem to be responsible for gradually increasing partial reflection of
later incoming waves, and this partial feed-back in turn enhances the
amplitudes of subsequent disturbances.  Growth is slow until the inner
disk reflects waves strongly enough to create a coherent linear
instability.

Note that the slow, fluctuating rise in $\delta_{\rm max}$ in both
runs 50M and 500M (upper panel of Figure~\ref{cntrst}) is roughly
exponential, with similar exponents in both cases.  This appears to
suggest that the behavior just described is, in fact, destabilizing in
the sense that it implies that the rate of growth is independent of
the amplitude.  It would seem, therefore, that no system of particles,
however large, could behave as a smooth disk.

\section{Summary and Discussion}
The main result presented here is the demonstration that scattering at
the inner Lindblad resonance of a spiral disturbance of even very low
amplitude causes a lasting change to the properties of the disk that
leads to increased amplitude of subsequent activity.  Linear
perturbation theory could never capture this behavior, because it
neglects second order changes to the equilibrium model caused by a
small amplitude disturbance.

The random density fluctuations of a system of self-gravitating
particles in a cool, shearing disk are amplified as they swing from
leading to trailing \citep{Toom90,TK91}.  The wave action created by
these collective responses is carried inwards at the group velocity to
the \ILR\ where it is absorbed by particles.  The resulting scattering
of particles to more eccentric orbits de-populates the near-circular
orbits over a narrow range of initial angular momentum.  Here I have
shown that, no matter how small the amplitude, the lasting changes to
the particle distribution promote a higher level of density
fluctuations that leads to indefinite growth.

The growth of non-axisymmetric waves in the idealized simulations
presented here is characterized by two phases.  When very large
numbers of particles are employed, the behavior appears to be that of
a system of dressed particles \citep{TK91,Wein98}, but one in which
the amplitudes of the particle wakes rise slowly.  As the peak
relative overdensity passes $\sim 2$\%, the changes to the
background state are destabilizing and simple instabilities appear
that exhibit run-away growth.

The instability that appears is caused by earlier changes to the \DF,
and is a true mode that could have grown exponentially from low
amplitude.  Direct tests reported in Section~\ref{restarts} demonstrate that
it does not rely upon non-linear coupling to previous structures.  I
speculate on a possible mechanism for the unstable mode in
Section~\ref{newmode}.

Linear perturbation theory \citep{Toom81} predicts no instability for
a smooth \DF, yet the evidence from Section~\ref{enhanced} is that no finite
number of particles would ever avoid indefinite growth.  Whether or
not this conclusion is correct, the avoidance of run-away growth for
some non-infinite number of particles is of theoretical interest only,
as galaxies are probably never as smooth as a disk of $5 \times 10^8$
randomly distributed particles that are each only $\sim 10$ times as
massive as a typical star.

Similar evolutionary changes to the original smooth disk are
undoubtedly the cause of the non-axisymmetric features that developed
for $m>2$ in the simulations reported by \citet{Sell11} to test the
models proposed by \citet{BLLT}.  The appearance of visible multi-arm
waves that heated the disk in those models was, as here, more and more
delayed as the number of particles was increased.  Note that no large
amplitude waves developed when forces were restricted to $m=2$ because
the disk was of such low mass.  The swing amplification of shot noise
is minimal when the parameter $X>3$, which is the case for bisymmetric
waves in this low-mass disk, and resonant scattering by such weak
disturbances will cause extremely slow growth, if any.  However,
$X\propto m^{-1}$, making the disk more responsive to disturbances
with $m>2$, allowing run-away growth of multi-arm features and disk
heating once these force terms were included.

\citet{TK91} used local simulations in a small shearing patch of a
disk to study the behavior of swing-amplified shot noise.  They were
able to explain even the long-term behavior with just linear theory,
although the limiting amplitude was a factor of $\sim 2$ higher than
they estimated, which they attributed to additional correlations
between the positions of particles that developed in the long-term.
Most significantly, they did not observe the kind of run-away growth
reported here.  This difference may be due to a combination of
factors: first, the particles in their principal box interacted with
pair-wise forces and, had the box extended infinitely in the azimuthal
direction, each non-axisymmetric feature would be composed of a
continuous spectrum of waves, implying that there would be no
equivalent of Lindblad resonances and the absorption of wave action
would be spread over a broad swath of particles.  However, their
periodic boundary conditions implied that the spectrum of waves was
discrete, and a Lindblad resonance must have been present for each of
the multiple waves that made up an individual shearing feature.  Thus
the absorption of wave action must have taken place at a number of
discrete locations, and mild resonant scattering probably occurred at
each.  But the most important distinguishing feature of their
simulations that prevented development of destabilizing features at
these resonances is that they applied a damping term to only the
radial velocity component of their particles.  Any resonant scattering
that depopulated nearly circular orbits must have been quickly erased,
preventing their system from manifesting the behavior described here.

I am grateful to the referee for suggesting I note the observations of
Saturn's rings \citep[reviewed in][]{Cuzz10}, which seem to support
the picture of spiral chaos proposed by \citet{Toom90}.  Of course,
the snowball particles in this beautiful ring system are believed to
collide dissipatively about once per orbit, which would again prevent
scattering at possible Lindblad resonances from creating reflecting
regions that might destabilize the disk.

While identifying the origin of indefinite growth is a good beginning
to a detailed understanding the origin of spiral features in $N$-body
simulations, it is far from being the whole story.  All the
simulations reported here supported several co-existing waves of
differing frequencies at all times, except runs 50Mc and 50Md in which
pre-existing waves were eliminated and single disturbance stood out
for a brief period.  The recurrence of large-amplitude waves remains
to be described in a later paper of this series.

In order to be able to reach the level of understanding described
here, I have simplified the dynamics of a realistic disk galaxy to the
extreme.  First, the particles in the simulations presented here
interacted with each other only through the $m=2$ sectoral harmonic of
the disturbance forces, with all other self-gravity terms being
ignored; even the mean central attraction was replaced by that of a
rigid mass distribution.  Second, the particles were constrained to
move in a plane only.  Third, I deliberately chose to study an
idealized model that was known \citep{Zang76,Toom81,ER98} to possess
no small-amplitude instabilities.  Fourth, I selected particles so
that their joint distribution of $E$ and $L_z$ matches that of the
adopted \DF\ as closely as possible.  Fifth, the simulations were
entirely isolated from external perturbations by passing satellites,
halo substructure, \etc, and the halo was represented by an idealized
axisymmetric and time independent mass distribution.  Sixth, all
effects of gas dynamics, such as the creation of dense clouds and the
formation of stars, were excluded.

Yet even with all these abstractions, the behavior took some effort to
understand.  The demonstration of a simple, clean instability was
possible only in restarted simulations, constructed by careful
reshuffling of the particles at the crucial time.  Simulations that
include all the above-mentioned complications continue to manifest
recurrent spiral patterns.  It is reasonable to hope that their origin
is related to the mechanism presented here, but this remains to be
shown.

The ultimate goal of this work is to demonstrate that some spirals in
galaxies are self-excited instabilities with a similar origin to that
in the simulations.  \citet{Sell10b} and \cite{Hahn11} have made a
step in this direction, by finding a feature in the distribution of
solar neighborhood stars that resembles that of scattering at an \ILR.
Further analysis of data from other ground- and space-based surveys,
such as {\small HERMES} \citep{BKK02}, {\small APOGEE} \citep{Maje10}
and especially {\it Gaia} \citep{Perr01}, will afford better tests.

\section*{Acknowledgments}
I thank Alar Toomre for providing the parameters of the linearly
stable model and for years of general encouragement to track down the
source of the spirals in simulations.  I also thank the referee for a
careful report, Scott Tremaine for some suggestions to improve the
presentation, and Tad Pryor for many helpful discussions.  This work
was supported by grants AST-0507323 and AST-1108977 from the NSF.

\label{lastpage}

\end{document}